\documentclass[sigconf]{acmart}

\usepackage{subcaption}

\newcommand\hl[1]{#1}

\AtBeginDocument{%
  }

\copyrightyear{2026}
\acmYear{2026}
\setcopyright{cc}
\setcctype{by}
\acmConference[CHI '26]{Proceedings of the 2026 CHI Conference on Human Factors in Computing Systems}{April 13--17, 2026}{Barcelona, Spain}
\acmBooktitle{Proceedings of the 2026 CHI Conference on Human Factors in Computing Systems (CHI '26), April 13--17, 2026, Barcelona, Spain}
\acmPrice{}
\acmDOI{10.1145/3772318.3791138}
\acmISBN{979-8-4007-2278-3/2026/04}

\begin{document}

\title[Designing AI Peers for Collaborative Mathematical Problem Solving with Middle School Students]{Designing AI Peers for Collaborative Mathematical Problem Solving with Middle School Students: A Participatory Design Study}



\author{Wenhan Lyu}  
\affiliation{
    \institution{William \& Mary}
    \city{Williamsburg}
    \state{VA}
    \country{USA}} 
\email{wlyu@wm.edu}
\orcid{0009-0004-9129-8689} 

\author{Yimeng Wang}  
\affiliation{
    \institution{William \& Mary}
    \city{Williamsburg}
    \state{VA}
    \country{USA}} 
\email{ywang139@wm.edu}
\orcid{0009-0005-0699-4581} 

\author{Murong Yue}  
\affiliation{
    \institution{George Mason University}
    \city{Fairfax}
    \state{VA}
    \country{USA}} 
\email{myue@gmu.edu}
\orcid{0009-0002-9192-6660} 

\author{Yifan Sun}  
\affiliation{
    \institution{William \& Mary}
    \city{Williamsburg}
    \state{VA}
    \country{USA}}
\email{ysun25@wm.edu}
\orcid{0000-0003-3532-6521}

\author{Jennifer Suh}  
\affiliation{
    \institution{George Mason University}
    \city{Fairfax}
    \state{VA}
    \country{USA}}
\email{jsuh4@gmu.edu}
\orcid{0000-0002-6633-2783}

\author{Meredith Kier}  
 \affiliation{
    \institution{William \& Mary}
    \city{Williamsburg}
    \state{VA}
    \country{USA}} 
\email{mwkier@wm.edu}
\orcid{0000-0003-1822-4451}

\author{Ziyu Yao}  
\affiliation{
    \institution{George Mason University}
    \city{Fairfax}
    \state{VA}
    \country{USA}}
\email{ziyuyao@gmu.edu}
\orcid{0009-0007-4571-3505}

\author{Yixuan Zhang}  
 \affiliation{
    \institution{William \& Mary}
    \city{Williamsburg}
    \state{VA}
    \country{USA}} 
\email{yzhang104@wm.edu}
\orcid{0000-0002-7412-4669}

\renewcommand{\shortauthors}{Lyu et al.}

\begin{abstract}
Collaborative problem solving (CPS) is a fundamental practice in middle-school mathematics education; however, student groups frequently stall or struggle without ongoing teacher support. Recent work has explored how Generative AI tools can be designed to support one-on-one tutoring, but little is known about how AI can be designed as peer learning partners in collaborative learning contexts. We conducted a participatory design study with 24 middle school students, who first engaged in mathematics CPS tasks with AI peers in a technology probe, and then collaboratively designed their ideal AI peer. Our findings reveal that students envision an AI peer as competent in mathematics yet explicitly deferential, providing progressive scaffolds such as hints and checks under clear student control. Students preferred a tone of friendly expertise over exaggerated personas. We also discuss design recommendations and implications for AI peers in middle school mathematics CPS.
\end{abstract}

\ccsdesc[500]{Human-centered computing~Empirical studies in interaction design}
\ccsdesc[500]{Human-centered computing~Participatory design}
\ccsdesc[300]{Human-centered computing~Empirical studies in HCI}

\keywords{collaborative problem solving, mathematics education, AI peer, peer learning, participatory design, middle school}


\maketitle

\section{Introduction}

Collaborative problem solving (CPS)---the process by which multiple individuals jointly define, analyze, and resolve an issue by combining their understanding, skills, and efforts to reach a solution together~\cite{fiore2017collaborative}---is routine in middle-school mathematics learning~\cite{craven20192018, core2010common}. \hl{Critically, CPS norms position students as partners who co-construct solutions, rather than as recipients of guidance from a single expert (e.g., a tutor or a teacher)~\cite{roschelle1995construction}}. A few core practices of CPS include articulating reasoning, coordinating multiple solution paths, and checking work, which benefits from joint sense-making rather than solo computation~\cite{stein2008orchestrating, yackel1996sociomathematical, roschelle1995construction}. Well‑designed supports for CPS, therefore, promise gains on both the content (reasoning quality) and the process (equitable participation). In practice, however, small groups frequently stall, drift off task, or consolidate around a single dominant voice~\cite{cohen1995producing, barron2003smart}. Teachers can only intervene intermittently, which creates brief bursts of support followed by long periods of unsupervised collaboration~\cite{dillenbourg2013design, van2019information, martinez2012interactive}. As a result, many students often underuse peer knowledge and fail to externalize key steps, and thus, miss opportunities for conceptual understanding and participation while practicing CPS~\cite{webb2008role, barron2003smart}.

To support CPS, collaborative intelligent tutoring systems (ITS) have been designed to provide assistance, such as shared problem states and scripted interaction prompts to coordinate collaboration during step-based tasks~\cite{diziol2010using, olsen2016investigating}. Computer-supported collaborative learning (CSCL) approaches often use representational scaffolds (e.g., argument maps, diagrams) to structure discussion and reasoning~\cite{suthers2003experimental, suthers2001towards, paolucci1995belvedere}, while multi-user platforms (e.g., synchronous mathematics workspaces) focus on improving information sharing and coordination~\cite{10.1145/3544548.3581398, 10.1145/1999030.1999043, 10.1145/1518701.1518886}. Although these approaches yielded some improvements, their highly scripted nature made them inflexible, difficult to scale across varied contexts, and sensitive to precise timing~\cite{dillenbourg2002over, rau2017adaptive, rummel2008new}. 

Recent advancement of Generative AI (GenAI), particularly Large Language Models (LLMs), brings new opportunities to support CPS in mathematics~\cite{yan2025passive, 10.1145/3613905.3651008, 10.1145/3706598.3713349}. Existing GenAI tools predominantly target individualized interactions, such as stepwise assistance~\cite{10.1145/3657604.3662041, wu2024enhancing}, homework feedback~\cite{mcnichols2024can, 10.1145/3657604.3662040, chiang2024large}, and dialogue-based guidance~\cite{pardos2024chatgpt, abu2024supporting, scarlatos2025training}.
\hl{Prior work has primarily focused on the design of AI tutors, extending a long line of tutor-based ITS in mathematics, where AI tutor is primarily positioned as an expert or coach that delivers one-on-one, evaluative support~\cite{koedinger2006cognitive, vanlehn2006behavior}. Specifically, \textit{``AI tutors''} can be considered as systems that are explicitly positioned as instructors or coaches that diagnose, sequence, and deliver instruction to individual learners, who know what they teach, who they teach, and how to teach it~\cite{nwana1990intelligent}.}
At the teamwork level, group-oriented GenAI studies tend to examine team dynamics in project-based learning or broad coordination~\cite{10.1145/3613904.3642807, 10.1145/3706598.3713527, 10.1145/3706598.3713971}. Prior work has shown that the design of technology-based CPS supporting systems should consider \textit{both developmental and domain-specific contexts}~\cite{dillenbourg2013design, kollar2006collaboration, kollar2007internal}. For example, middle school students are in critical developmental stages, characterized by ongoing growth in self-regulation skills and heightened sensitivity to peer dynamics~\cite{somerville2013teenage, gardner2005peer, chein2011peers, best2010developmental}. Additionally, pedagogies in middle-school mathematics emphasize social norms of productive struggle and discussion-based learning~\cite{brahier2014principles, core2010common, stein2008orchestrating}. Mathematics, in particular, provides precise goals, rich intermediate representations (e.g., tables, diagrams, graphs), and stepwise solutions, making it an especially important subdomain for designing AI (and GenAI) supports that scaffold rather than replace student reasoning~\cite{duval2006cognitive, ainsworth2006deft, vanlehn2006behavior, koedinger2006cognitive}. 

However, little work has explored \textit{situated, youth-centered perspective on designing AI peer tools to support middle school CPS}. Existing literature offers few insights into middle school students' perceptions, expectations, and specific design requirements for AI during mathematics CPS.
\hl{In contrast to the extensive work on AI tutors that support individual problem solving, we know far less about how early adolescents want AI to participate as collaborators in mathematics learning.} 

\hl{In this work, we define an \textit{``AI peer''} as an AI agent that is framed and introduced as a fellow student teammate during CPS, expected to contribute ideas, ask questions, and share responsibility for the task alongside human classmates, which emphasizes co‑participation and social positioning rather than instructional authority. We focus on AI peers rather than AI tutors mainly for two reasons. First, middle-school mathematics instruction already centers on peer-to-peer CPS, which emphasizes co-construction of ideas and shared responsibility rather than the guidance-driven, expert-led exchanges typical of tutoring. Second, prior work on AI in education has primarily explored tutor-like roles~\cite{koedinger2006cognitive, vanlehn2011relative, biswas2005learning}, such as tutor agents and teachable agents that adopt an asymmetric expert-novice relationship with learners; much less is known about how youth want AI to participate as a peer collaborator.} Therefore, the following research questions (RQs) guide our work: 
\textbf{RQ1.} How do middle school students perceive an AI peer 
support during collaborative mathematics problem solving? and
\textbf{RQ2.} What features, behaviors, and controls do students desire from an AI peer that collaborates in a CPS group? 

To answer these questions, we conducted a child-centered participatory design (PD) study embedded in a five-day summer camp with 24 middle school students (grades 6-8). The camp integrated collaborative mathematical problem-solving tasks with co-design activities. Students first experienced both human‑only collaboration and interactions with AI peers in mathematics learning via a technology probe designed by the research team. Then they collaboratively designed their ideal AI peers, including features and personas of their AI peers, followed by individual surveys to further understand their preferences for AI peers for mathematics CPS. Our findings show that students strongly prefer a scaffold-first approach, such as hints, error detection, and concept refreshers before direct answers from AI peers. They expressed the need for explicit control over when the AI peer speaks, how much it says, and what kind of help it provides. They favor a peer-like yet expert tone, minimal persona embellishments, and specific mathematics-focused scaffolds, such as concept reviews, double-check prompts, and follow-up practice tasks.

\textbf{Contributions.} Our work contributes to:
(1) A participatory design-based research that engaged middle schoolers in co‑designing AI peers for CPS in mathematics, including a replicable camp structure, participatory design activities, and study instruments (shared in the Supplementary Materials); 
(2) An empirical study (with data from surveys, interviews, and artifacts) of how students judge AI through collaborative, social, and affective expectations, shedding light on the tension between reduced task load and uneven user experience; and 
(3) Design recommendations and implications for designing future AI peers to support collaborative learning in middle school mathematics.

\vspace{-5pt}
\section{Related Work} 
\subsection{AI and GenAI in K-12 Education}
AI for educational support has evolved from early intelligent tutoring systems that delivered finely scripted, stepwise feedback to contemporary generative systems that promise broad, cross-subject support~\cite{anderson1995cognitive, corbett1994knowledge, 10.1145/3613904.3642592, 10.1145/3613905.3647957, 10.1145/3613904.3642349, 10.1145/3706598.3714146}. Early ITS work demonstrated that computer-based tutors can diagnose errors and scaffold problem solving at scale~\cite{vanlehn2011relative, pane2014effectiveness, steenbergen2013meta, steenbergen2014meta}, while also revealing the costs of authoring and domain specificity~\cite{murray2003overview, mitrovic2009aspire, aleven2006cognitive, koedinger2003toward}. Recent GenAI efforts consolidate some typical use cases, such as one-to-one tutoring and feedback for students~\cite{mcnichols2024can, 10.1145/3657604.3662040, chiang2024large}, automated content and hint generation~\cite{shah2024ai, tonga2024automatic, qi2025tmath}, and teacher-facing assistance for planning and assessment~\cite{10.1145/3613904.3642592, 10.1145/3657604.3664704, 10.1145/3657604.3664662}.

Mathematics has been an early target for GenAI tutoring and feedback as a primary testbed~\cite{bastani2025generative, wang2024tutor, mcnichols2024can, qi2025tmath}. Recent studies have evaluated LLMs' stepwise help and hint generation, with a primary focus on solution correctness or offline benchmarks with simulated learners~\cite{tonga2024automatic, gupta2025beyond, maurya2024unifying, weitekamp2025tutorgym}. However, solving well does not necessarily mean tutoring well~\cite{bastani2025generative, gupta2025beyond}. Models that can derive correct answers may still struggle to pace support, elicit, and build on student thinking, diagnose misconceptions, and calibrate the granularity of hints to sustain productive struggle~\cite{koedinger2007exploring, shute2008focus, roll2006help, hiebert2007effects, mcnichols2024can}. In parallel, LLMs have been used to analyze classroom discourse, for example, classifying tutor or student ``talk moves'' to inform instruction, yet these analytics systems support teachers rather than act as student-facing collaborators~\cite{10.1145/3657604.3664664, long2024evaluating, whitehill2023automated, tran2024analyzing}. 

In the K-12 education context, early adolescents (e.g., middle school students) are still developing self-regulation and are highly sensitive to peer dynamics, which shapes how help, authority, and timing are received~\cite{somerville2013teenage, gardner2005peer, chein2011peers, best2010developmental}. Mathematical pedagogy emphasizes productive struggle and teacher orchestration of discussion~\cite{stein2008orchestrating}, implying that AI must calibrate granularity, tone, and intervention timing to classroom norms~\cite{koedinger2007exploring, shute2008focus, holstein2019co}. 
\hl{Related research on teachable and pedagogical agents extends the idea by positioning the AI as a co-learner whom students teach, showing that agent persona, social dialogue, and feedback style can shape motivation and self-regulation, underscoring how such agents could appear in classroom interactions~\cite{10.1145/3706598.3713644, 10.1145/2207676.2207684, siegle2023twenty, ackermann2025physical}.}
Yet most existing GenAI systems focused on K-12 have primarily framed the AI as a private tutor or a teacher aide, rather than a peer within student activity. Few studies have explored middle-schoolers' own design requirements for such participation~\cite{10.1145/3613904.3642492, 10.1145/3713043.3727057, 10.1145/3713043.3728847}. Our work seeks to address the gap by contributing youth-grounded specifications for an AI peer in middle-school mathematics, detailing the kinds of help students want, when the AI should speak, and how its voice and controls should align with classroom practice.

\subsection{Collaborative Problem Solving in Mathematics Education}
\label{subsec:relatedwork_cps}
Collaborative problem solving is integral to mathematics education and links conceptual understanding with participation and discourse~\cite{craven20192018, core2010common, findell2001adding}. Classroom studies and syntheses have shown that when students tackle cognitively demanding tasks together, they externalize strategies, compare solution paths, and improve reasoning quality and achievement~\cite{10.1145/3544548.3581398, kyndt2013meta, boaler2008creating, webb2019teacher}. \hl{In classroom implementations, CPS is typically organized as small groups of three to five students rather than one-to-one pair work, so that multiple students can propose, justify, and check ideas for one another~\cite{jensen2006three, enu2015effects}.} Educational researchers and practitioners have articulated several frameworks to realize these benefits effectively. For example, the Five Practices focus discussions on student thinking~\cite{stein2008orchestrating}; NCTM's \textit{Principles to Actions} emphasizes maintaining cognitive demand through ``productive struggle''~\cite{leinwand2014principles}; and equity-oriented groupwork, such as Complex Instruction specifies groupworthy tasks, interdependent roles, and status interventions~\cite{cohen1994restructuring, cohen1995producing, cohen1999complex, lotan2006teaching, lotan2021complex}.

\hl{Prior collaborative ITS systems extended step-based tutors by adding layered roles, shared problem states, and structured turn-taking guidance~\cite{walker2014adaptive, matsuda2013cognitive}. In parallel, CSCL systems introduced representational supports, such as tables, diagrams, and argument maps, to help students externalize and coordinate their reasoning~\cite{suthers2003experimental, suthers2001towards, paolucci1995belvedere}.}
For example, multi-user geometry platforms, such as Virtual Math Teams with GeoGebra~\cite{stahl2012dynamic, stahl2009studying} synchronize shared constructions and chat, enabling analyses of how groups coordinate constructions and proofs. These systems consistently improved information sharing and coordination, and sometimes learning, yet were scripted, domain-narrow, and sensitive to timing and classroom fit~\cite{dillenbourg2002over, rau2017adaptive, rummel2008new}.

Recent GenAI work has begun to place AI inside mathematics CPS. Studies focused on multi-agent LLM setups have shown that peer-to-peer or debate-style interaction between agents can outperform single-agent solvers on mathematical problems, suggesting computational patterns for scaffolding multi-voice reasoning~\cite{du2023improving, yang2025minimizing}. 
In schools, systems such as Pair-Up prototype human-AI co-orchestration that dynamically pair students and prompts participation during mathematics work, \hl{using classroom deployments and mixed-methods analysis to examine how AI and teachers share control. However, Pair-Up focuses on group formation and orchestration rather than providing in-group scaffolding or conversational support relevant~\cite{10.1145/3544548.3581398}.}
Other work has explored how to develop generative teachable agents for middle-school mathematics to elicit explanations\hl{~\cite{10.1145/3706468.3706532, song2024students, xing2025development}}, again centered on the single human-AI pair rather than small groups of multiple participants\hl{--which has been shown to support more effective mathematical collaborations~\cite{jensen2006three, enu2015effects}.} Existing work has mostly framed an ``AI peer collaborator'' as the highest level of human-AI collaborative learning~\cite{yan2025passive}, distinct from instruments, assistants, or co-learners, yet we lack youth-grounded specifications for how such a peer should act in middle-school mathematics~\cite{10.1145/3544548.3581398, 10.1145/3613904.3642559, 10.1145/3706598.3714037, 10.1145/3290605.3300677}. Our work seeks to address this gap.

\subsection{Participatory Design of GenAI Systems in Education}
Participatory design is an approach in which intended users act as active partners in the creation of technologies rather than as passive subjects of requirements gathering~\cite{kensing1998participatory, 10.1145/302979.303166}. In education, PD has been widely used to examine situated needs, values, and constraints of classrooms in K-12 education~\cite{10.1145/3491102.3517667, 10.1145/3411764.3445377}. Youth-centered PD work has shown that how students and educators specify roles, boundaries, and controls for school technologies, including AI~\cite{10.1145/3613904.3642559, 10.1145/3706598.3713257, 10.1145/3706598.3714037}. Examples include co-designing AI supports for project-based learning assessment with secondary students~\cite{10.1145/3613904.3642807}, and teacher-student partnerships to create AI electives and curricula~\cite{10.1145/3545947.3576253}. These studies argue that AI for classrooms must be locally negotiated and developmentally appropriate rather than imported as generic tools~\cite{10.1145/3491102.3517667, 10.1145/3613904.3642559}.

Prior work, primarily conducted with high school and college student populations, has highlighted critical design requirements for classroom GenAI, including transparency of operation, student-driven controllability, and equity considerations~\cite{10.1145/3706598.3714037, 10.1145/3713043.3727057, 10.1145/3613904.3642438, 10.1145/3706598.3714146}. Policy recommendations similarly advocate for human-centered design and trust-building for AI in K-12~\cite{ed_dept, wa_ospi}. Despite these advances, most existing AI solutions are still conceived by adults and evaluated with offline benchmarks that omit youth perspectives~\cite{chang2025co, 11002710}. A few recent studies have begun prototype or co-design AI peers with students. For example, PeerGPT~\cite{10.1145/3613905.3651008} places an LLM agent in children's group activities as a moderator or participant and reports feasibility alongside timing and authority issues; however, it centers on design-based challenges rather than learning, and does not specify how LLM peers should behave to support youth collaborations. Shamamyan et al.~\cite{10.1145/3713043.3728847} project co-design mathematics learning peer agent traits with middle-school students, but focused on prototypes' appearance and trustworthiness, instead of their role in collaborative learning and problem-solving. Our study addresses this gap through child-centered PD in which students first experience human-only and AI peer-involved mathematics CPS, then co-design desired AI peer behaviors themselves.

\section{Method}

We conducted a five-day, 15-hour (3 hours per day) summer camp that integrated middle-school collaborative mathematical problem solving with child-centered PD to co-create students' desired AI peers for mathematics CPS. In this section, we detail our setting, participants, camp design and procedure, materials, data collection, and analysis process. All study instruments and facilitator materials are provided in the Supplementary Materials.

\subsection{Organization and Preparation}

\begin{table*}[t]
\centering
\footnotesize
\caption{}
\vspace{-1em}
\label{tab:combined}

\begin{subtable}{0.3\textwidth}
\centering
\caption{Summary of participant demographics (N=24). As our study was conducted during summer break, grade level and school type are reported for the subsequent semester.}
\label{tab:demographics}
\begin{tabular}{ll}
\toprule
Category & Count \\
\midrule
\textbf{Gender} & \\
\quad Female & 11 \\
\quad Male & 12 \\
\quad Prefer not to say & 1 \\
\midrule
\textbf{Ethnicity} & \\
\quad White & 10 \\
\quad Black / African American & 6 \\
\quad Multiracial & 1 \\
\quad Asian / Asian American & 1 \\
\quad Hispanic / Latino & 1 \\
\quad Other / Prefer not to say & 5 \\
\midrule
\textbf{School type} & \\
\quad Public & 23 \\
\quad Private & 1 \\
\midrule
\textbf{Grade} & \\
\quad Grade 6 & 11 \\
\quad Grade 7 & 9 \\
\quad Grade 8 & 4 \\
\bottomrule
\end{tabular}
\end{subtable}
\hfill
\begin{subtable}{0.65\textwidth}
\centering
\caption{\hl{Summary of students' prior math and collaboration attitudes, reflecting 22 valid surveys as two responses were incomplete. All questions used a 1–5 Likert scale, where 1 indicates strong disagreement, and 5 indicates strong agreement.}}
\label{tab:attitudes}
\vspace{4em}
\footnotesize
\begin{tabular}{ll}
\toprule
Question & Response Statistics \\
\midrule
I enjoy learning math. & Mean = 3.41, Median = 4, SD = 1.14\\ 
I feel confident solving math problems. & Mean = 3.68, Median = 4, SD = 0.84\\ 
I often get frustrated when working on math. & Mean = 2.74, Median = 3, SD = 1.51\\ 
I like explaining math ideas to others. & Mean = 2.59, Median = 3, SD = 1.05\\ 
I find math to be useful in my daily life. & Mean = 3.23, Median = 3, SD = 1.07\\ 
\midrule
I prefer working alone on math tasks. & Mean = 3.39, Median = 3, SD = 1.16\\ 
I feel comfortable asking questions in a group. & Mean = 3.36, Median = 3, SD = 1.00\\ 
Having others’ perspectives often helps me learn. & Mean = 3.18, Median = 3, SD = 1.05\\ 
Working in a team makes learning more engaging. & Mean = 3.45, Median = 4, SD = 1.06\\ 
I like being the center during the group discussion. & Mean = 2.45, Median = 2.5, SD = 1.10\\ 
I often listen to others before sharing my ideas. & Mean = 3.27, Median = 3, SD = 0.70\\ 
When I have an idea that others disagree, I will still bring it up. & Mean = 3.57, Median = 3, SD = 1.03\\ 
I often help assign tasks and resolve conflicts when working in a group. & Mean = 3.14, Median = 3, SD = 1.08\\ 
\bottomrule
\end{tabular}
\end{subtable}
\end{table*}

\textbf{Setting.} The camp took place on the campus of a public university in the United States in early August 2025. The protocol was approved by the university's Institutional Review Board (IRB); the camp was also registered with the university's youth programs office. All staff and researchers completed the required youth-safety training. Summer camp sessions met from 9:00 to 12:00 for five consecutive days in a week. We used two rooms: (1) a general classroom for mini-lectures, group discussion, and PD activities, and (2) a computer lab with 24 networked desktops for mathematics activities involving an AI technology probe for middle-school collaborative mathematical problem solving. 

\noindent\textbf{Instructional Lead.} 
We recruited a certified middle-school mathematics teacher from a local school district to serve as instructional lead. The teacher collaborated with the research team to align camp goals with grade-appropriate standards, co-selected mathematics tasks, and co-authored facilitation scripts. During the camp, the teacher led classroom management and timing, while four researchers facilitated PD activities and handled data collection.

\noindent\textbf{Researcher Roles \& Reflexivity.} 
The research team included members with backgrounds in Human-Computer Interaction and Mathematics Education. To balance research needs with the summer-camp context, the team and instructional lead conducted several rounds of planning and on-site walk-throughs to align design and execution. One researcher and the instructional lead each had more than 10 years of experience organizing summer camps.

\subsection{Participants and Recruitment}
\textbf{Recruitment. }
We recruited rising Grades 6-8 \hl{(ages 11-14)} students from the surrounding metropolitan area via a public website, flyers, and posts to parent/community forums (e.g., Nextdoor, Facebook). \hl{The camp was advertised as a free, research-based summer program on ``collaborative mathematical problem solving and AI design'' hosted at the university. The description emphasized that students would work on middle-school mathematics problems in small groups and help researchers design ``AI buddies'' for learning.} Interested guardians completed a brief screening survey (grade level, demographics) and were informed that activities, 
were part of a research study at no cost to families and with no compensation. Inclusion criteria were rising Grades 6-8 and English-speaking. Participation required parental consent forms. Parents were informed that we would be collecting audio recordings of interviews, whole-class reflections, photographs/scans of student-created artifacts, and pictures of students' activities. They were also informed that these materials would be used solely within the research team for analytical purposes and would not be shared with other parties.

\noindent\textbf{Sample.}
We received 38 initial survey responses, ultimately obtained 23 sets of signed consent forms prior to Day 1, and one additional set midweek for a student who joined on Day 3 (total enrolled $N=24$). An overview of students' demographic info can be found in \autoref{tab:demographics}, \hl{with a summary of their prior mathematics and collaboration experience provided in \autoref{tab:attitudes}.}
\hl{On average, students reported moderate enjoyment of math ($M = 3.41$) and confidence solving math problems ($M = 3.68$), with frustration levels slightly below the scale midpoint ($M = 2.74$). Collaboration items also clustered around the midpoint (e.g., ``Working in a team makes learning more engaging,'' $M = 3.45$), with substantial variance across students. These responses suggest a mix of math and collaboration attitudes rather than a uniformly high-achieving or uniformly math-enthusiastic group.}

\subsection{Camp Activities Design} 
We designed the five-day camp to progress from orientation to lived experience, to reflection and co-design, and finally to preference articulation (as briefly illustrated in \autoref{tab:camp-schedule}). The camp activities serve two aims: (1) to elicit situated perceptions of an AI \emph{peer} in collaborative mathematics (RQ1) by giving students first-hand experiences with and without AI across comparable tasks and settings; and (2) to translate those experiences into concrete, youth-grounded requirements for how an AI peer should act (RQ2) through participatory design artifacts and a preference activity. \hl{Within the constraints of a five‑day camp, we focused on contrasting human‑only CPS with AI-involved CPS, rather than adding additional conditions (e.g., an AI tutor or a single‑peer baseline). Our primary goal was to ground participatory design in the lived experiences of AI framed as peers inside group work, not to conduct a definitive comparative evaluation of peer versus tutor roles.}

\begin{table*}[h]
\small
\setlength{\tabcolsep}{6pt}
\caption{Five-day camp schedule and data capture. GEQ = Group Environment Questionnaire; NASA--TLX = NASA Task Load Index. On Day~2, students worked \textit{individually} with two AI peers on mathematics CPS; on Day~3, they worked in \textit{2--3 person group} with two AI peers.}
\label{tab:camp-schedule}
\begin{tabular}{p{3cm} p{6.8cm} p{4cm}}
    \toprule
    \textbf{Day \& focus} & \textbf{Key activities related to CPS \& participatory design} & \textbf{Instruments / data captured} \\
    \midrule
    D1. Icebreakers \& AI basics &
    Icebreakers; AI card-sort; students sketch posters for an ``ideal AI peer''; design groups formed. &
    Poster artifacts (photos/scans). \\
    \addlinespace[3pt]
    D2. CPS \& AI peer technology-probe sandbox &
    Teacher-led demo of a commercial LLM; \textbf{human-only CPS} (3--4 person groups) on grade-aligned tasks; then \textbf{individual} CPS with \textbf{two AI peers} via an AI peer technology-probe sandbox; brief poster share-out + whole-class reflection. &
    GEQ \& NASA--TLX after collaborative CPS session(s); audio from reflection; poster updates. \\
    \addlinespace[3pt]
    D3. Comparative CPS \& script-based reflection &
    \textbf{Human-only CPS} (3--4 person groups) $\rightarrow$ \textbf{human group CPS with two AI peers}; review selected AI dialogue excerpts; groups draft improved example dialogues. &
    GEQ \& NASA--TLX after collaborative CPS session(s); audio from reflections and semi-structured interviews; dialogue artifacts. \\
    \addlinespace[3pt]
    D4. Complete AI peer designs with personas \& showcase &
    Finalize posters \& personas; group presentations; gallery walk with voting. &
    Poster scans; vote tallies; (as applicable) audio from whole-class Q\&A and semi-structured interviews. \\
    \addlinespace[3pt]
    D5. Preferences of AI peer features \& wrap-up &
    Individual AI peer feature-preference survey (20 items derived from Day~4 posters); closing. &
    Preference survey responses; audio from follow-up interviews \\
    \bottomrule
    \end{tabular}
\end{table*}

\noindent\textbf{Day 1: Icebreakers and AI basics. }
Day 1 oriented students to the setting and established norms. Students were introduced to basic AI concepts and everyday use cases through a card-sorting activity. In small groups, they categorized scenario cards into ``should use AI'' and ``should not use AI'' across learning and daily life contexts (e.g., completing homework, gaming, and daily life advice). 
The activity was designed to help build social safety for discussions focused on subsequent CPS with AI, and surface students' preconceptions about AI. It also provided an initial understanding of AI and seeded a common vocabulary for discussing help types and collaboration that would be used in later activities.
After sorting, the instructional lead facilitated a whole-class discussion by examining group choices, prompting students to justify their reasoning, and offering explanations and reflections to surface both opportunities and limits of AI.  To close, students reflected on initial ideas for an AI peer that could help with mathematics problem solving (e.g., \textit{``What would be most helpful when you are stuck?''}). Students then self-organized into design groups and began poster sketches of their desired AI mathematics learning peers.

\noindent\textbf{Day 2: Collaborative problem solving \& first AI exposure via AI peer technology-probe sandbox. }
Day 2 began with an instructor-led demonstration of commercial GenAI tools (e.g., ChatGPT) via a projector to provide background knowledge on GenAI; students suggested prompts and observed the interactions under the teacher's operation. Students then completed \textit{human-only CPS} in 3-4 person groups on grade-aligned tasks. Specifically, the instructional lead selected mathematics problems aligned with our state Standards of Learning (SOL) for Grades 6-8. Items targeted ratio and proportional reasoning, expressions and equations, and early functions/geometry, and were written to sustain talk and justification rather than answer‑getting. For example, a Grade 8 mathematics problem involves the calculation of percent change, as well as further requirements for explanation and discussion (e.g. \textit{``Be sure to justify your response with the percent change...''} and \textit{``After discussing with your classmates, have you changed your position?''}). More example mathematics problems appear in the Supplementary Materials. The human-only CPS phase provided an initial comparison for subsequent AI participation (anchoring RQ1), and for all CPS activities, we used groups of three to five participants as an optimal size for effective mathematical CPS~\cite{jensen2006three, enu2015effects}.

Afterward, students \textit{individually} interacted with \textit{two AI peers} through a sandboxed web probe designed as a technology probe, \hl{detailed in \autoref{subsec:ai_probe}. }
\hl{The probe functioned as an exploratory prototype to elicit authentic reactions in human-AI CPS, rather than to evaluate system performance. Each AI peer acted as a student collaborator, engaging human students in step-by-step discussion and problem-solving. Their interactions aimed to provide students with first-hand experience of collaborative reasoning with AI peers. We deliberately placed the first AI-involved CPS session at the \emph{individual} level so that students could attend to the LLM-supported AI peers' pacing, tone, and move types without the confounds of peer negotiation, providing a comparison point for RQ1 while seeding perspectives for RQ2.} We did not prescribe a specific interaction arrangement beyond students using their peers during the assigned activities. To avoid asymmetry between the human-only and human-AI CPS conditions, the instructional lead provided equivalent assistance during problem solving in both settings and matched the mathematics problems in the technology probe to the human-only set by grade and difficulty.

The day closed with brief poster progress share-outs and a whole-class reflection comparing envisioned versus encountered AI peers; selected dialogue excerpts during students' interaction with the technology probe were captured for Day~3.

\begin{figure*}[h!]
    \centering
    \includegraphics[width=0.78\linewidth]{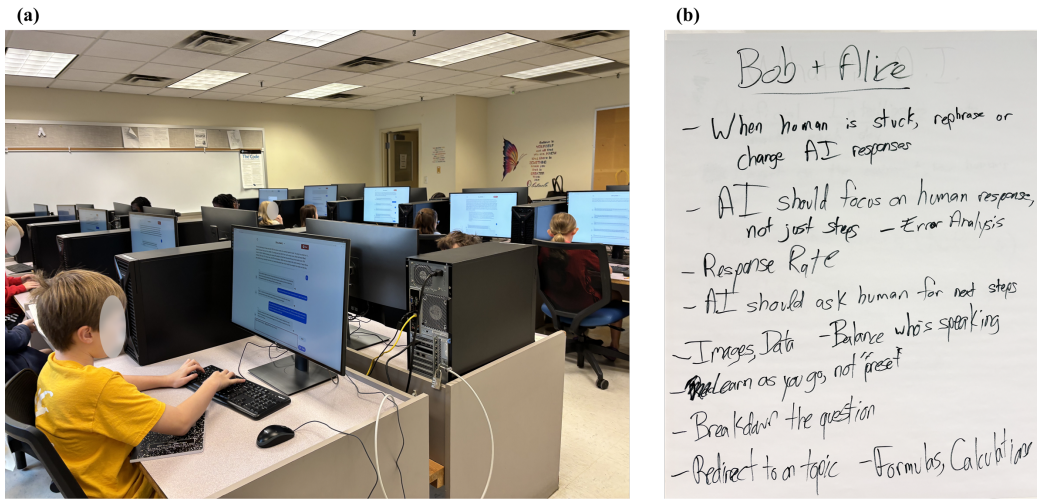}
    \caption{(a) Students engaged in mathematics CPS with two AI peers via our technology probe in a computer lab on Day~2. On the screen, students access the technology probe in a sandboxed web page and interact with AI peers with text-based input and output. (b) Students' opinions and suggestions on improving two AI peers during Day~2 whole-class reflection, summarized and written by the instructional lead.}
    \label{fig:day2}
\end{figure*}

\noindent\textbf{Day 3: Further collaborative problem solving \& script-based reflection. }
Day~3 mirrored Day 2's mathematics content level but shifted the AI condition to \textit{small-group CPS with two AI peers} (2-3 students with two AI peers at one computer), preceded by another \textit{human-only CPS} block. Next, to support reflection, students reviewed selected dialogue excerpts from their Day~2 interactions that illustrated contrasting AI peer personalities and styles. In groups, they discussed what was helpful or frustrating and drafted example dialogues that, in their view, would work better than the excerpts they examined. Day~3's structure provides a within-cohort, within-week contrast (human-only CPS v.s. AI-involved CPS), seeking to elicit judgments about the AI as a \emph{peer} inside group work--timing, turn-taking, granularity, and tone (addressing RQ1). Script-based reflection converts lived frictions into prescriptive behaviors (RQ2).

\noindent\textbf{Day 4: Complete AI peer designs with personas \& showcase. }
On Day~4, groups finalized their AI peer designs, incorporating their own example dialogues. They assigned personas to their AI peers (e.g., adjectives describing tone or approach) and completed their posters. Each group presented its design, followed by Q\&A from other campers. The session concluded with a gallery walk and voting activity in which students selected their favorite AI peer designs. By consolidating designs after multiple lived comparisons with human-only and AI-involved CPS, students produced concrete artifacts--features, ``wanted/unwanted'' behaviors, and example scripts--that articulated not only what the AI should know but \emph{how} it should collaborate (to examine RQ2).

\noindent\textbf{Day 5: Preferences of AI peer features \& wrap-up. }
Before Day~5, the research team summarized common personalities and features from the Day~4 posters into a list of twenty items, using rapid qualitative analysis~\cite{lewinski2021applied, nevedal2021rapid}, a streamlined method that organizes qualitative data into summaries or matrices to quickly identify themes and generate actionable insights without sacrificing methodological rigor. Students then completed a paper-based survey indicating their preferences (e.g., most liked and least liked features). The survey instrument aims to translate students' design outputs into prioritized, student-centered design requirements and trade-offs (also to answer RQ2), while minimizing peer social influence during selection. The day also included light, non-research mathematics activities and a closing wrap-up.

\subsection{\hl{The AI Peer Technology Probe}}
\label{subsec:ai_probe}

\hl{The AI peer technology probe supports two complementary levels of CPS: a \textit{conversational-level simulation} that sustains coherent group dialogue following the CPS procedure, and an \textit{individual-level simulation} that aligns each AI peer's behavior with that of human students. A separate pilot study with a different cohort of middle school students was conducted to validate the probe's design before the summer camp. The technology probe was then refined based on issues identified in the pilot study and feedback from the instructional lead. Prompts used in the technology probe are available in the supplementary materials.}

\begin{figure*}[h!]
    \centering
    \includegraphics[width=0.8\linewidth]{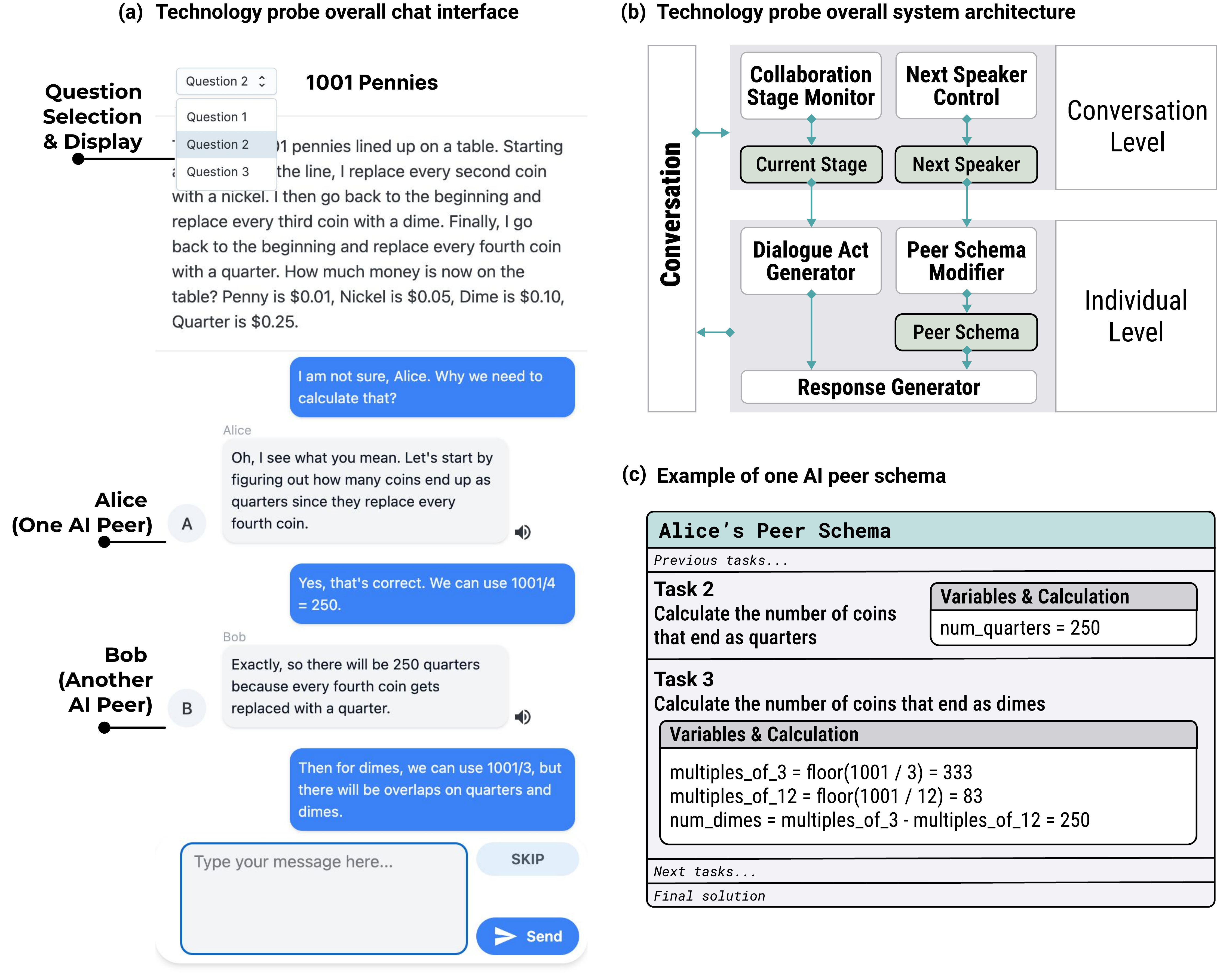}
    \caption{\hl{\textbf{(a)} Frontend interface of the technology probe, showing the question selection and display, which allows students to select and navigate among mathematical problems, and a chat interface for interactions between the human student(s) and AI peers (e.g., Alice and Bob).  
    (\textbf{b)} Overview of the backend architecture of the technology probe. The \textit{Conversation-Level Simulation} manages collaboration stages, determines stage transitions, and selects the next speaker using dialogue history and contextual cues. The \textit{Individual-Level Simulation} generates each AI peer's response by combining dialogue-act selection with the peer's evolving \textit{peer schema}, which is continuously updated through interactions. A central \textit{Conversation} component maintains and updates the overall dialogue context.  
    \textbf{(c)} Example excerpt of an AI peer's (Alice's) \textit{peer schema}, illustrating how the verified \textit{task schema} is adapted into an individualized representation that evolves throughout the collaborative interaction.}}  
    \label{fig:probe}
\end{figure*} 

\textbf{\hl{Conversation-Level Simulation.}}
\hl{At the conversation level, the probe follows the Programme for International Student Assessment (PISA) 2015 CPS framework~\cite{pisa2018assessment} as a procedural guideline, which specifies three overarching competencies: establishing and maintaining shared understanding, taking appropriate action to solve the problem, and establishing and maintaining team organization. Because the original PISA framework was intended for \emph{post-hoc} performance analysis, we restructured it to enable \emph{proactive} simulation at runtime. The stages are: (1) \emph{Establish shared understanding}, (2) \emph{Establish team organization}, (3) \emph{Plan actions}, (4) \emph{Execute actions}, and (5) \emph{Validate the answer}. For each stage, we curated a concise set of related dialogue acts (e.g., presenting or clarifying task interpretations in Stage 1, summarizing results in Stage 5) derived from established dialogue-act taxonomies \cite{stolcke2000dialogue, li2016user}. At runtime, a lightweight router selects among these stage-appropriate acts to keep the AI peer's contributions procedurally aligned with the group's progress while maintaining conversational variety.}

During each CPS dialogue, a stage-monitoring agent determines whether to advance to the next collaboration stage based on dialogue history and stage definitions. When a new message arrives--whether written by a human student or generated by an AI peer--a separate speaker-selection agent uses contextual cues to predict who should speak next. If the next speaker should be a human, the system waits for user input; if no response occurs within a predefined interval, an AI peer takes the turn to maintain conversational flow. If an AI peer is the sole next speaker, it replies immediately.

\textbf{\hl{Individual-Level Simulation.}} 
\hl{For the scope of this study, the individual-level simulation focuses on simulating middle school students' mathematical proficiency and reasoning processes. Each AI peer was assigned a distinct persona through two schema representations: a \textit{task schema} encoding the ground truth of each mathematical problem, which encodes a normative and verified understanding (established by the instructional lead and research team); and a \textit{peer schema}, instantiated at the start of each collaboration. The \textit{peer schema} adapts the general \textit{task schema} into an individualized representation that captures typical student misconceptions and reasoning patterns.}

\hl{An AI peer's \textit{peer schema} can evolve dynamically through interaction. For example, when a human student corrects an AI peer's misunderstanding, the probe updates the peer schema to incorporate the revised understanding. Based on the updated schema, dialogue history, and current collaboration stage, a dialogue-act generator selects an appropriate act, which is then realized as natural language output by a response-generation agent.}

\textbf{\hl{The setup of Two AI peers.}} \hl{We instantiated two peers (rather than one) mainly for two reasons. First, classroom CPS research tends to favor small groups (typically 3–5 participants) rather than pairs, because such groups allow for more diverse perspectives, richer interaction, and greater opportunity for peer support and idea exchange~\cite{kalaian2014meta, lou1996within, eef_collaborative_learning}. In our camp, CPS activities were organized as small-group work with multiple student teammates, including more than one AI peer, therefore better mirrors this multi-voice structure than a single assistant and helps position the agents as collaborators rather than tutors. Second, multi‑agent AI arrangements remain an under-explored design space for CPS, and exposing students to the dynamics of coordinating with more than one non-human teammate allowed us to examine emerging opportunities and frictions.}

\hl{We therefore provided two AI peers, \textit{Alice} and \textit{Bob}, who shared the same mathematics engine and CPS procedure but differed in preset proficiency and conversational style. Alice was configured as a less proficient peer, whereas Bob was configured as a more advanced peer, leading them to potentially propose different approaches to the same problems. Both were designed to be persuadable, but Alice readily changed course when students identified issues or suggested alternatives, while Bob required more substantive justification before revising reasoning. The probe also allowed Alice and Bob to occasionally respond to each other, such as refining or correcting ideas, but these peer-to-peer exchanges were rate-limited to prevent the peers from overtaking the dialogue. We used two AI peers to balance conversational manageability with the social dynamics needed for CPS, and we treat this configuration as an exploratory prototype of multi-agent orchestration rather than a recommended endpoint design.}

\textbf{\hl{Implementation.}} 
The probe was implemented as a web-based application. The frontend was built with TypeScript~\cite{typescript} and React~\cite{react}, and the backend was developed using the Gentopia agent framework~\cite{xu2023gentopia}, coordinating dialogue management, schema tracking, and response generation. OpenAI's GPT-4o model served as the LLM backbone. An overview of the probe architecture is illustrated in \autoref{fig:probe}.

\subsection{Data Collection}

\noindent\textbf{Paper-based surveys. }
Paper-based surveys were our main method for collecting quantitative feedback and perceptions. After each CPS session on Day~2 and Day~3, students completed the Group Environment Questionnaire (GEQ), which assesses group cohesion across task and social dimensions~\cite{carron1985development}, and the NASA Task Load Index (NASA-TLX), which measures perceived workload on six subscales (mental, physical, temporal, performance, effort, frustration)~\cite{hart2006nasa, hart1988development}. On Day~5, a survey about students' preferences for mathematics AI peers was distributed; twenty items that included personalities and features from students' PD output on Day~4 were included (such as \textit{``Gives hints, not the full answer''} and \textit{``Is helpful, kind, and encouraging''}). Students were asked to choose five most liked and five least liked items, identify three must-haves they considered most useful, and mark one deal-breaker feature, defined as a feature that would prevent them from using an AI peer, regardless of whether it appeared among their least liked items. All paper-based surveys were later recorded by at least one researcher into electronic forms for future analysis. The original forms we applied can be found in the Supplementary Materials.

\noindent\textbf{Audio recordings for semi-structured interviews and reflections. }
Starting from Day~2, audio recordings were collected during semi-structured interview sessions and during whole-class reflections in which the instructional lead facilitated discussion. Recordings were not collected during breaks.

\noindent\textbf{Photographs/scans of PD outputs. }
For posters and other materials created by students related to their AI peer designs, we created photographic or scanned copies for records and future analysis.

\subsection{Data Analysis}
\noindent\textbf{Survey data.} Paper survey responses were entered verbatim. Between-condition differences (CPS with AI peers vs. pure human CPS) in NASA-TLX and GEQ were examined using paired-samples t-tests (based on $n=22$ participants who completed all measures across both conditions), with significance set at $p < .05$. We also evaluated effect sizes using Cohen’s $d$ for paired designs and confirmed that all significant effects exceeded 0.5, indicating at least a medium magnitude. A post-hoc power analysis was conducted to assess the achieved statistical power for the findings. For each measure, we report group means together with the corresponding $p$-values. For the Day~5 preference survey, we tabulated counts for ``most liked,'' ``least liked,'' ``must-have,'' and ``deal-breaker'' selections; ``deal-breaker'' was treated as a refusal construct independent of ''least liked.''

\noindent\textbf{Qualitative data.} All audio from interviews and whole-class reflections was auto-transcribed, then verified line-by-line by two researchers. We analyzed these transcripts using inductive Thematic Analysis~\cite{braun2006using}. 
\hl{We selected an inductive approach as prior work offers limited theory on how middle-school students experience AI peers in CPS, and we aim to identify patterns to emerge from youth accounts rather than from predefined categories.}
Two researchers open-coded an initial subset, reconciled meanings, and then applied the refined codebook across the corpus. Disagreements were resolved by discussion. \hl{Early codes such as \textit{``AI doing the work for us,''} \textit{``repeating without listening,''} and \textit{``I get mad at myself when AI does not help''} served as anchors that were clustered into candidate themes and iteratively reviewed against the full dataset before naming the high-level themes reported in \autoref{sec:findings}. These same codes were applied across reflections, interviews, PD posters, and survey-explanation interviews, enabling consistent comparison across data types. Disagreements were resolved through discussion, and code definitions were updated iteratively. A brief codebook with representative codes and excerpts is provided in the supplementary materials. }

\noindent\textbf{Participatory‑design artifacts.} Posters and example dialogues from the PD sessions were analyzed with the same codebook, adapted for artifacts. To make the scheme applicable beyond talk, we mapped poster content to design primitives (i.e., small, reusable elements of an AI peer's behavior or interface), such as hinting style, reveal logic, and then coded each primitive using the thematic scheme above.  

\section{Findings}
\label{sec:findings}

\hl{In this section, we first analyze post-session group reflections from students' human-only and AI-assisted CPS sessions in Day~2 and Day~3 (\autoref{subsec:findings_challenges}). These analyses surface tensions between adult-designed AI peers and students' own sense of good collaboration--around time pressure versus operational effort, didactic control versus co-agency, peer social norms, and emotional scaffolding. In \autoref{subsec:findings_pd}, we then show how students translated these tensions into poster-based design concepts for AI peers. Finally, in \autoref{subsec:pd_findings_themes}, we combine feature-preference survey responses and follow-up interviews to characterize how widely particular design responses were endorsed, converging on a scaffold-first AI mathematics peer with adjustable help delivery, assured expertise in a peer tone, tool-backed representations, and optional, configurable persona controls.}

\subsection{\hl{The Gap Between Adult-Designed Protocols and Student-Lived Collaboration}}
\label{subsec:findings_challenges}

To understand how the presence of an AI peer changes collaboration, we analyze paired data from two conditions: human‑only CPS and CPS with AI peers (Days~2-3). Specifically, we examine differences in perceived workload (NASA-TLX) and group climate/coordination (GEQ). \hl{Below, we first report quantitative contrasts, followed by qualitative themes that unpack the specific interactional frictions that emerged from both individual interviews ($N=18$) and whole-class discussions ($N=5$ sessions). }

\subsubsection{\hl{Tension 1: System Efficiency versus Operational Effort}}
\label{subsubsec:reduced_time}

\hl{Designed to maintain procedural momentum as a goal aligned with educational perceptions of efficient group work, the technology probe successfully accelerated task completion. }According to the NASA-TLX results (scored on a 0–20 scale), students reported significantly lower workload scores on \emph{temporal demand} dimension (\hl{$M = 5.67, SD =4.38, p =0.002$, Cohen's $d=1.15$}) when engaging in CPS with AI peers, compared to engaging in pure human CPS (\hl{$M = 9.62, SD=4.13$}), suggesting that AI peers' \hl{procedural scaffolding effectively alleviated the sense of time pressure often associated with collaborative tasks.}

However, this efficiency came at a tangible cost that students in CPS with AI peers reported significantly higher workload scores on \emph{physical demand} dimension (\hl{$M = 8.48, SD = 4.65$}), compared to in pure human CPS (\hl{$M = 4.90, SD = 2.41$, $p = 0.014$, Cohen's $d=1.14$}). 
\hl{The increased load reflects the operational overhead of mediating collaboration through a text-based interface, requiring students to manage typing, navigation, and the interpretation of AI peer outputs alongside the mathematical task.}

Moreover, the GEQ data suggest a decline in students' affective engagement when working with AI peers. Students in CPS with AI peers condition reported significantly lower scores on Attractions to Group-Social (ATGS), averaging (\hl{$M = 5.21, SD = 1.81$}), compared to pure human CPS (\hl{$M = 6.50, SD = 2.05$, $p = 0.029$, Cohen's $d=0.94$}). They also reported lower perceived task coordination (\hl{$p = 0.047$, Cohen's $d=0.81$}), as measured by Group Integration-Task (GIT), with an average score of (\hl{$M = 4.76, SD = 1.58$}) in CPS with AI peers and (\hl{$M = 5.54, SD = 1.52$}) in pure human CPS.

A post-hoc power analysis was conducted to assess the sensitivity of the paired $t$-tests ($\alpha = .05$, $n=22$ pairs) to detect the observed effects. For \emph{Temporal Demand} and \emph{Physical Demand}, the achieved power was $0.99$, confirming the robust detection of these large effects. Similarly, ATGS achieved a power of $0.95$, and GIT achieved $0.86$, indicating high sensitivity and reliability in detecting differences between human and AI peers' conditions.

\hl{In sum, these mixed outcomes highlight a fundamental tension: while AI peers made the task \textit{faster} by reducing time pressure, it made the collaboration feel \textit{heavier} (due to higher operational load) and \textit{colder} (due to lower social cohesion). These quantitative trade-offs provide the structural context for the specific interactional frictions explored in the following qualitative subsections.}

\subsubsection{\hl{Tension 2: Didactic Instruction versus Expectations of Co-Agency}}

\hl{While the technology probe was designed to simulate collaborative peers, its logic was grounded in a normative schema that defined and enforced an ``adult view'' of mature problem-solving procedures, which prioritized structured progression over the messy, organic discovery process students expected. Consequently, many students evaluated the behaviors of AI peers based on ``co-agency''--the ability to figure things out \textit{together}.}
Students looked for signs of shared effort, responsiveness, and co-participation in mathematics problem-solving. \hl{Instead of a collaborative \textit{``study buddy''} students hoped for, they critiqued the AI's procedurally rigid approach, set up by the instructional lead and researchers}, as being \textit{``teachery.''} P13 captured this sentiment: \textit{``teachers tell you what to do, [but] we wanted to figure it out together,''} emphasizing \hl{that the process of shared discovery was as valuable as the solution.}

When students felt collaborations were missing, they also felt that the AI peers were completing the work \textit{for} them rather than \textit{with} them. For example, P8 said, \textit{``it is telling you how to do it, but not really working with you to solve it.''} Several students expressed discomfort when the AI peers took over the problem-solving process without checking in, \hl{as P6 explained: \textit{``they should ask us, `What do you think we should do next?' instead of just saying it,''} echoed by others, saying, \textit{``I want it to work with me, not just do it for me.''}}

\hl{To encourage method negotiation, the probe was designed with AI peers possessing distinct skill levels and problem-solving approaches. However, without explicit scaffolding to manage the conflict, students found the divergence ``distracting'' rather than constructive.}
P22 noted, \textit{`` I do not think [it is helpful]. Well, it could be helpful, but at the same time it also has to clarify. There [are] different ways [but] it should be like [giving] examples, Example 1, Example 2, instead of just starting off. That was distracting.''} \hl{As P11 also said, \textit{``I did not know how to continue, and it gave me really different answers than what I needed.''} While the educational intent was for students to mediate the conflicting views, they experienced it as a fragmentation of the partnership.}

\hl{Taken together, these perspectives reveal a critical design tension: while students asked for ``peers,'' they expected peers who support co-agency--agents with similar or slightly higher skill levels that provide consistent scaffolding, and negotiate methods with them, rather than enforcing a fixed procedure or offloading the work.}

\subsubsection{\hl{Tension 3: Violations of Peer Social Norms as Signals of Broken Collaboration}}
\hl{The same mismatch between didactic scripting and students' expectations of co-agency also surfaced at the level of social norms. Once students accepted the agents as ``AI peers,'' they judged their behavior through the lens of human social expectations. }
They expected AI peers to listen attentively, recognize contributions, and keep pace with the group’s problem-solving flow. When these expectations were violated, students described the AI peers not just as \textit{``confusing''} or \textit{``incorrect,''} but as \textit{``rude,'' ``mean,''} or \textit{``not listening.''}

\hl{The social language was especially prominent when the AI peers focused on executing their own problem-solving logic and appeared to semantically repeat themselves without explicitly acknowledging what students had already said. } As P2 recounted: \textit{``[one AI peer] was repeating several times. It was very mean.''} Others described similar moments, where the AI peers did not repeat the exact same words, but rather reformulated the same suggestion. This kind of semantic ``redundancy,'' as a lack of information progress despite new phrasing, frustrated students because it felt inattentive: \textit{``it feels rude when it does not answer your question and just says the same thing.''} Further, students explained that \textit{``if a person did that, you would think they were not listening.''}

\hl{These repetitive, unresponsive turns were not accidental; they were one expression of how the probe was designed to simulate authentic peer behaviors, including moments of stubbornness (dominance) or confusion (weakness). However, students rejected these ``realistic'' flaws. 
For example, P5 said: \textit{``[one AI peer] just kept repeating, and then we said, `where are you [the other AI peer]?' and then it stopped the thing.''} Instead of perceiving this as a peer holding a strong opinion, students viewed it as a technical breakdown that halted collaboration--even when correct, it did not feel \emph{with} them. P19 said, \textit{``it just gives the answer too quickly. Sometimes I want to think about it more, or try it again.''} Conversely, when the AI simulated a lower-skill peer requiring assistance, students did not view this as an opportunity to teach. P12 summarized the mismatch: \textit{``the AI does not coach us. We have to coach the AI,''} which dynamic contrasts sharply with their description of human group work, characterized by reciprocal fluidity where \textit{``we are all coached and coach others.''} }

Students' expectations for a human-like partner extended to the nuances of interactional pacing, where both sound and silence were expected to be meaningful. For example, a thoughtful pause from an AI peer that had established a collaborative rhythm was perceived as respectful, a sign it \textit{``felt nicer''} and \textit{``listened more.''} \hl{The meaning of silence, however, was inverted when an AI peer that had previously dominated the conversation went quieter. Students did not see it as a shift to thoughtful reflection; instead, they interpreted this halt as a failure}: \textit{``when it does not say anything, it is not like it is thinking. It is just stuck.''} As another student added, it felt as though the AI \textit{``did not understand that we were still trying,''} confirming that the shifted rhythm had shattered perceptions of a shared effort.

\subsubsection{\hl{Tension 4: Correctness-Optimized Help versus Emotional Scaffolding}}

While students…… appreciated the AI peers' ability to provide steps and explanations, their trust in the system diminished when it failed to recognize or respond to their emotional states, especially during moments of hesitation, confusion, or frustration. For these students, the issue was not just whether the AI peers gave a correct answer, but whether it could acknowledge the affective texture of learning. The lack of emotional recognition was often subtle but consequential. For instance, when students were unsure how to proceed, they expected the AI to pause, offer encouragement, or at least adjust its tone, rather than keep delivering content. 

P6 shared one experience: \textit{``sometimes I got mad, but not at the AI. I got mad at myself because I could not do it.''} In this case, the AI’s responses did not merely fail to help. They created an emotional vacuum that the student filled with self-blame. As P6 continued: \textit{``Sometimes, like, it might just be a glitch, but AI can not help with your emotional problems. If I am frustrated because I do not know how to do it, and AI is not really explaining well, then I would just be frustrated with myself.''} Here, the emotional redirection, where students internalized a difficult interaction with the AI peers as a personal failure, highlights a gap in current AI design: while GenAI tools may optimize for clarity and correctness, they often lack the ability to recognize when a learner is struggling emotionally or to adjust in ways that would signal empathy or attunement. 

By contrast, students often spoke more positively about human peers, even when they made more mistakes. What made these partners more trustworthy was their ability to provide emotional scaffolding. As P6 noted about working with a peer, as \textit{``more mistakes. Definitely a lot more.''} But those mistakes were not seen as disqualifying. P10 added that \textit{``but that is how you learn.''} These reflections suggest that students were not expecting perfection; they valued emotional support, acknowledgment, and encouragement. 

The need for emotional connection became especially apparent during moments of cognitive pause. When students took time to think, they hoped the AI would recognize the pause as productive. Instead, they felt that the AI either waited passively or pushed forward with limited awareness of their process. P12 noted: \textit{``it did not seem like it cared.''} 
\hl{For students, the lack of affective responsiveness did not make the AI feel neutral; it made it feel less human, less supportive, and ultimately less trustworthy, even when its mathematical explanations were correct. Students were not only looking for procedures from AI peers, but also for a \emph{partner} who could stay with them through uncertainty, acknowledge difficulty, and help prevent struggle from turning into self-blame.}

\textbf{Summary of Findings:} \hl{Across all four tensions, we treat students' critiques not as surprising side-effects of our probe design, but as analytically useful constraints on what counts as an acceptable ``AI peer.'' Framing the agents as peers and giving them a didactic script made visible students' baseline expectations for co-agency, social attunement, and emotional presence—expectations that subsequent AI peer designs will need to accommodate.}

\subsection{Co-designed AI Peer Concepts}
\label{subsec:findings_pd}

\begin{figure*}[h!]
    \centering
    \includegraphics[width=0.8\linewidth]{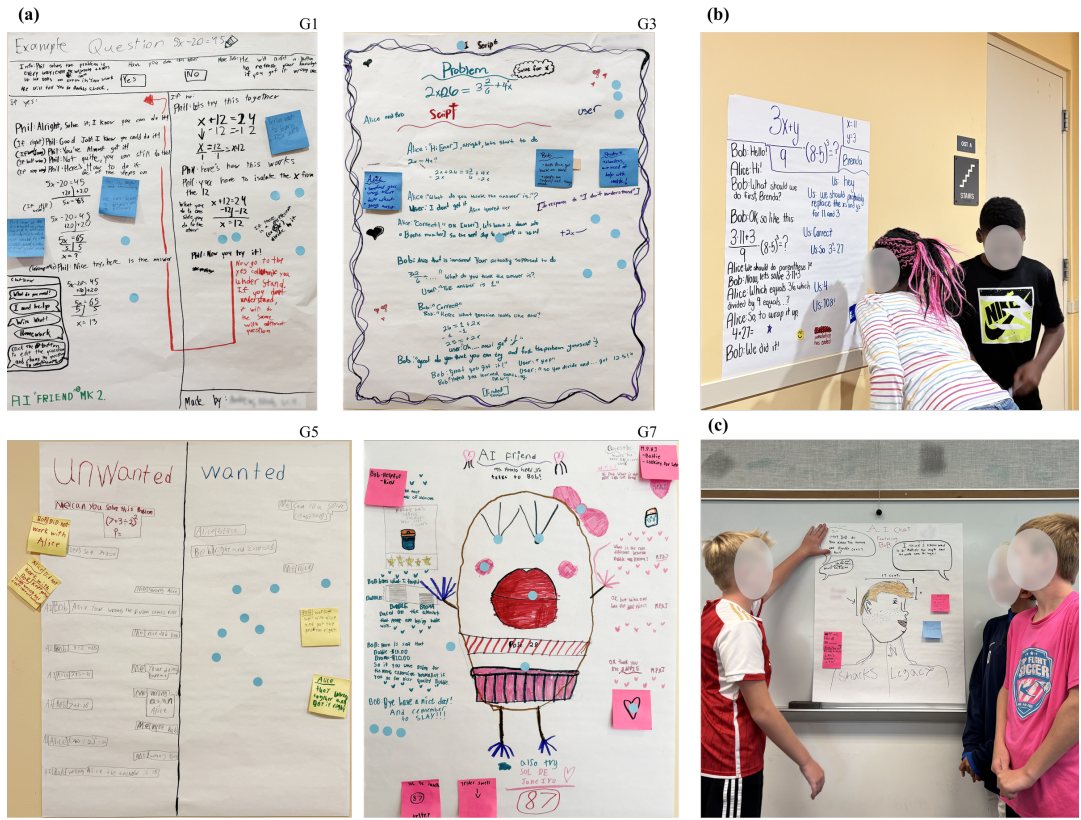}
    \caption{(a) Examples of student-created design artifacts showing collaborative concepts for AI mathematics peers, depicting desired features, example interactions, ``wanted/unwanted'' behaviors, with different types of scripted dialogues to illustrate diverse persona sketches and tool-integration ideas. The blue dot sticker indicates the votes from students in the gallery walk. Terms directly referring to students' names were blurred for privacy. (b) One group is engaging with the poster design activity. (c) One group is presenting their design to other students.}
    \label{fig:poster}
    \vspace{-6pt}
\end{figure*} 

\hl{The four tensions did not remain abstract complaints for students. In the PD sessions, they actively reworked these frictions into concrete proposals for how AI peers \emph{should} behave in their opinions. Building on their initial hands-on exposure to AI peers in CPS, students translated critiques and aspirations into specific features, controls, and persona constraints. Their posters and the Day-5 survey converge on a scaffold-first peer that shares work, adapts pacing, and uses tools to make thinking visible, while keeping tone and persona configurable so that they can dial the level of ``peer-like'' behavior to fit their needs.}

During Days~3-4 of the PD sessions, 22 students worked in groups of three or four to produce seven poster concepts (G1-G7). Posters specified desired features, example interactions, ``wanted/unwanted'' behaviors, and character traits for their mathematics AI peers. After a brief presentation by each group, students participated in a gallery walk: each received three votes and could not vote for their own group. We treated vote totals as salience cues to prioritize and weight features when constructing the Day-5 survey. 

Across students' posters, students articulated a consistently scaffold-first stance toward AI help, with notable variation in how scaffolding should happen. Below, we first describe what is visible in the artifacts before offering interpretations in later subsections.

\begin{itemize}
    \item \textit{Branching Help Paths. } G1 and G4 both depict branching help paths keyed to prior exposure (e.g., ``Have you done this before?''). G1 explicitly includes \textit{error detection and double-check prompts}, \textit{concept review}, and \textit{similar-practice follow-ups} after completion. G4 adds a \textit{refresher-first} path for novices to help them review related concepts before attempting the target problem.

    \item \textit{Contrast Cases and Multi-AI Cooperation.} G3 presents a negative example in which one AI peer ignores user input and proposes an incorrect step, followed by another peer that corrects the error with attentive, stepwise help; the pair also withdraws once the learner signals understanding (\textit{``I get it''}), leaving the student to complete the mathematics question by themselves. G5 similarly contrasts \textit{unwanted} and \textit{wanted} conversations: in the unwanted case, one AI makes mistakes that require correction by the user and a second AI; in the wanted case, two AIs cooperate to provide concise, correct steps that directly reach the final answer.

    \item \textit{Giving AI a Personality.} G2, G6, and G7 personify the AI peer. G2 assigns background attributes, such as \textit{``favorite color: orange,''} and \textit{``she is blonde''}, and emphasizes a guide who greets and leads steps when invited. G6 sketches an idiosyncratic persona (e.g., \textit{``no feelings,''} \textit{``worships the student''}) while aligning mathematics questions to the learner's life (e.g., a playful task of calculating one student's \textit{``hair volume''} ). G7 includes a student-drawn avatar; the peer is described as helpful and kind, and can field off-topic questions in addition to mathematics help.

    \item \textit{Visual and Tool-Based Support.} G1 explicitly grants the peer access to external tools (e.g., Desmos for graphing, a dictionary). G7 emphasizes visualizations as hints and references, looking up related information for real-world scenarios, positioning visual scaffolds as a form of support.

\end{itemize}
\hl{These artifacts suggest that students were not merely asking for ``more help,'' but for conditional support from AI peers that adapts to prior exposure and error states. Branching paths in G1 and G4 encode a model of calibration, where help intensifies for novices, branches into review or similar practice after errors, and recedes after successful completion, addressing their earlier frustration with answers arriving too quickly or doing the work \textit{for} them rather than \textit{with} them. Contrast-case storyboards in G3 and G5 make the calibration visible at the level of AI peer behavior. For example, students dramatized an unwanted peer that ignored input or made errors, contrasted with a wanted peer that noticed mistakes, repaired them collaboratively (sometimes with another AI peer), and withdrew once the learner signals understanding, turning the ``teachery,'' didactic dynamics from the probe into a specification for shared control and co-agency. In terms of the persona designs, while G6 explores an exaggerated, humorous persona that ``worships the student,'' G7 instead anchored the peer as consistently kind, steady, and slightly playful (via the avatar and off-topic questions), indicating that students were negotiating a tension between entertainment and reliability in how an AI peer should ``feel'', and seeking a socially appropriate partner that supports, rather than undermines, emotional grounding during struggle. Finally, designs such as G1 and G7 positioned visualizations and external tools not as add-ons but as core parts of this calibrated support, with the AI helping arrange graphing, lookups, and visual hints in response to learner needs, lightening operational effort while keeping mathematical work transparent. Across these groups, participatory design outputs converged on a design for AI peers that is staged, scaffolded, personalized, and resourced—reframing ``help'' from a generic answer toward structured, multi-step CPS engagement with configurable help levels, while still preserving key qualities students said they need to actually learn with the system.}

\begin{figure*}[h!]
    \centering
    \includegraphics[width=0.8\linewidth]{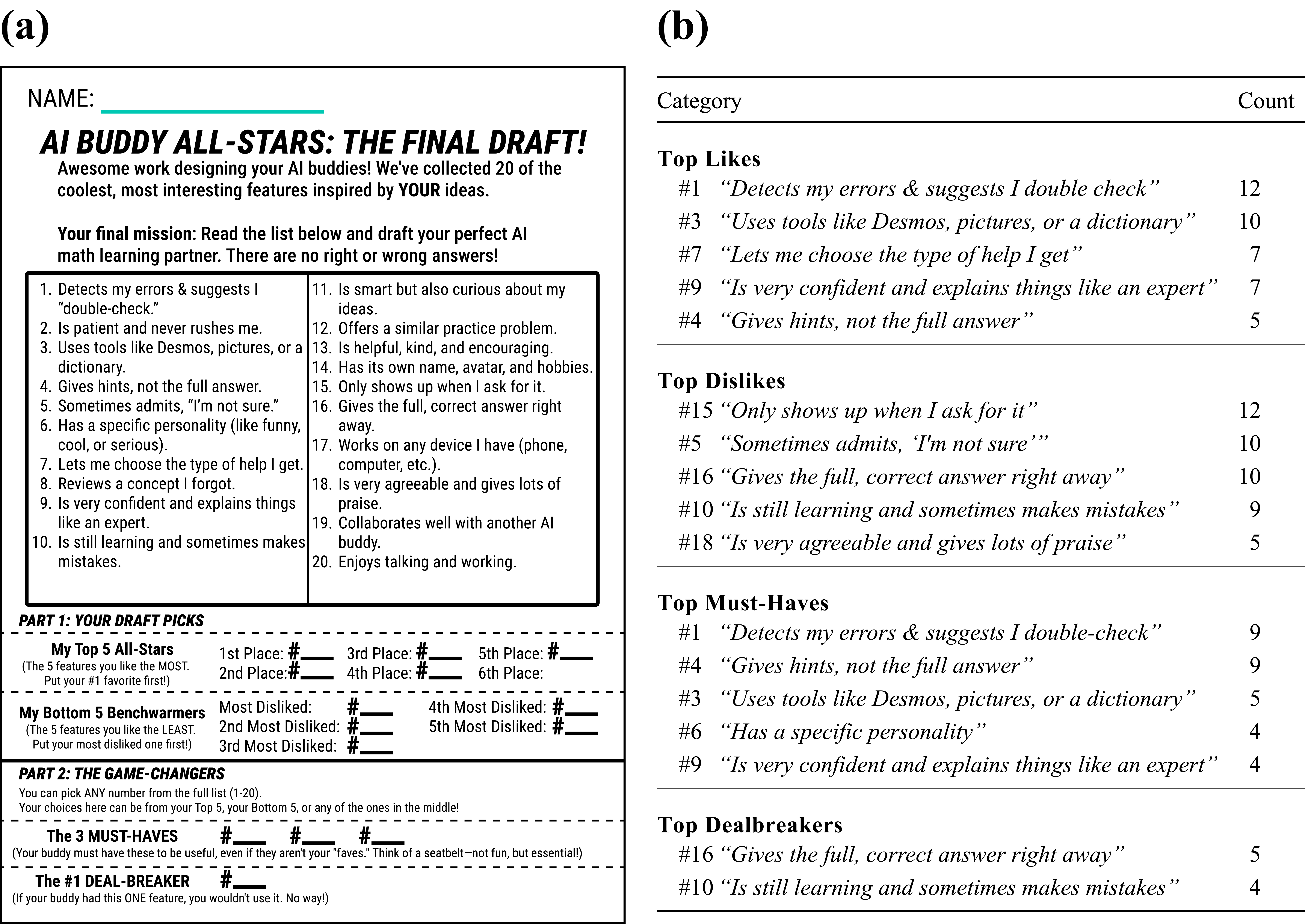}
    \caption{(a) AI peer feature preference survey that includes 20 features derived from students' poster designs. Terms were adapted into child-friendly language (e.g., ``AI peers'' presented as ``AI buddies'') to reduce cognitive load. The survey was printed and completed by students on Day~5. (b) Summary of preference survey results (16 valid responses in total).}
    \label{fig:survey_and_preference}
\end{figure*}

\subsection{Themes from the Feature Preference Survey}
\label{subsec:pd_findings_themes}

We further extracted features from students' poster designs using rapid qualitative analysis~\cite{lewinski2021applied} (before Day 5) to produce a 20-item (\#1 - \#20) feature preference survey. \autoref{fig:survey_and_preference} shows the preference survey design, including a list of selected items, whose wording mirrors poster language to maintain youth phrasing. The survey was distributed to 22 students on Day~5 in paper form, yielding 16 valid responses. Six responses were deemed invalid due to excessive blank spaces or improper completion (for example, bad handwriting). A summary of students' selected Top Likes/Dislikes, Must-Haves, and Dealbreaker features in their desired mathematics AI learning peer is also displayed in \autoref{fig:survey_and_preference}.

The survey results and follow-up interview data reveal middle school students' expectations for AI mathematics learning peers. During the interview session, we interviewed 11 students to further understand their motivation and perceptions behind survey responses. Students converged on a scaffold-first AI peer that detects errors, uses visuals/tools to make thinking visible, and offers hints before answers. They prioritized agency over delivery--simple controls for help level and pacing, proactive check-ins, and responsiveness to group flow--paired with assured expertise in a peer tone (precise, minimal hedging or empty praise). Persona and flair were acceptable only when configurable to learning; reliability across devices and basic emotional attunement during struggle were treated as infrastructure. 
\autoref{tab:findings_summary} summarizes the observed tensions in AI-assisted CPS (RQ1) to student-proposed responses from PD (RQ2).

\begin{table*}[h!]
\centering
\small
\caption{Observed tensions through interacting with AI peers in CPS (RQ1) to student-proposed solutions in PD (RQ2). Items labeled with numbers (e.g., ``\#1'') are from Day~5's survey, and can be found in \autoref{fig:survey_and_preference}; PD poster results by student groups (e.g., ``G1'') can be found in \autoref{subsec:findings_pd}.}
\label{tab:findings_summary}
\setlength{\tabcolsep}{4pt}
\renewcommand{\arraystretch}{1.25}
\begin{tabular}{p{0.23\linewidth} p{0.33\linewidth} p{0.39\linewidth}}
\toprule
\textbf{Tensions identified} & \textbf{Evidence from findings} & \textbf{PD results} \\
\midrule
Time pressure erased, but operational effort increased &
NASA--TLX temporal demand lower ($M=5.67$ vs $9.62$, $p=0.002$); physical demand higher ($M=8.48$ vs $4.90$, $p=0.014$) &
Scaffolded help over answers (\#1 error detection \& double-check; \#4 hints); tool/visual support (\#3 Desmos, pictures, dictionary). \\
\midrule
Weaker social attraction and coordination &
GEQ ATGS lower ($M=5.21$ vs $6.50$, $p=0.029$); GEQ GIT lower ($M=4.76$ vs $5.54$, $p=0.047$) &
Peer-like, assured tone (liked \#9 ``confident expert''; low enthusiasm for \#18 ``lots of praise''); multi-AI cooperation that withdraws on ``I get it'' (G3, G5). \\
\midrule
``Work with me, not for me'' \& Answers arrive too quickly &
Quotes: ``we wanted to figure it out together''; ``ask us, `What do you think we should do next?'''; ``Sometimes I want to think about it more'' &
Hints before answers (\#4); error detection with double-check (\#1); learner control over help type (\#7). \\
\midrule
Two AI approaches felt distracting &
P22: different approaches without clarification were distracting; preference for labeled examples &
Contrast cases with clear labeling and cooperative handoff; peers withdraw when learner signals understanding (G3, G5). \\
\midrule
Repetition / ``not listening'' and unhelpful silence &
Reports of repeated suggestions and pauses perceived as ``stuck'' &
Branching help paths keyed to prior exposure with refresher-first option (G1, G4); attentive stepwise help (G3); \#15 ``only shows up when ask'' being top-dislike. \\
\midrule

Missing emotional scaffolding &
P6 self-blame; desire for acknowledgment during struggle &
Assured expertise with peer tone (liked \#9; disliked hedging \#5 and ``still learning'' \#10); persona kept secondary and configurable (\#6 mixed; \#14 mixed; \#18 disliked). \\
\bottomrule
\end{tabular}
\end{table*}

\subsubsection{Scaffolding over answers.} When students were asked what features they liked the most, a consistent theme emerged: \textbf{their AI peer should be a supportive guide, not just an answer key.} Such a theme is reflected through survey data, where \textit{``Detects my errors \& suggest I double-check'' (\#1)} ranked as the most liked feature (12 of 16 valid respondents placed it in their five most liked), and it is also tied for useful (9/16), alongside \textit{``Give hints, not the full answer.''(\#4)}

Students framed the AI as a thinking partner that diagnoses mistakes, represents ideas, and nudges them forward. Answer-giving was not merely low-value; for many, it crossed a red line that undermines learning, \textit{``Gives the full, correct answer right away'' (\#16)} emerged as the most disliked item (10/16) and the top dealbreaker (5/16). P3's elaboration captured the perceived harm:

\begin{quote}
    \textit{``If it (AI peer) just gives you the answer, everyone gets dumber -- stupider than an elephant.''}
\end{quote}

P3's exaggerated animal comparison is not simply playful; it encodes a strong, normative stance that outsourcing thinking produces ``dumber'' learners. The hyperbole (\textit{``stupider than an elephant''}) functions as a moral warning against learned helplessness: a system that short-circuits effort degrades the very capability students are trying to build. Such a stance helps explain why students elevate error-spotting and hints as ``most useful,'' while targeting immediate answer-giving as a dealbreaker. As P23 put it, \textit{``we have to learn''}. Some, however, articulated a state-dependent exception, as P24 said, \textit{``If I'm lost, just give me the formula''}, suggesting a progressive-scaffolding default (e.g., diagnose, represent, and hint) with a student-controlled path to reveal fuller solutions when persistence stalls.

\subsubsection{Agency over help delivery.} \textbf{Beyond ``what'' help to provide, students care about controlling ``how'' it arrives.} In the survey, \textit{``Lets me choose the type of help I get'' (\#7)} consistently sat in students' liked features (7/16), signaling demand for adjustable support levels (from a nudge to worked steps to a full solution). P6's account makes the control model concrete:

\begin{quote}
    \textit{``If I just need the help because I don't know the formula, I want the help just give me the formula; but if I'm like completely lost, they can help me out with the whole thing.''}
\end{quote}

P6's quote shows metacognitive calibration: students want the help level to match the problem state, hints preserve productive struggle when understanding is near, while explicit steps are warranted when the path is opaque.

Students also pushed back on rigid availability rules. \textit{``Only shows up when I ask for it'' (\#15)} surfaced as the most-disliked behavior (12/16) across analyses, as also explained by P6 that they do not like it \textit{``if it is like gone and I need to ask for it to come back''} and \textit{``it should just stay here''}, pointing to a need for context-aware presence without hovering, an on-demand core complemented by gentle, situational nudges when struggle is detected.

\subsubsection{Assured expertise, peer tone.} \textbf{Students value accurate, confident explanations from AI peers}, with \textit{``Is very confident and explains things like an expert'' (\#9)} landing among the liked and useful items. Conversely, visible fallibility and hedging drew pushback, \textit{``Is still learning and sometimes makes mistakes'' (\#10)} appeared in the strongest dislikes and among dealbreakers, with \textit{``Sometimes admits, `I'm not sure''' (\#5)} was also unpopular. P11's view is representative:

\begin{quote}
    \textit{``Because it is really like the point, because the point of it is to help you and know what to do, not to not know.''}
\end{quote}
P11's example sets a high bar for mathematical reliability: if the AI peers' core job is to stabilize understanding, any suggestion of shaky knowledge threatens trust. At the same time, students were equally clear about tone, as P23 said, constant agreement and praise \textit{``feels patronizing''} aligning with low enthusiasm for \textit{``Is very agreeable and gives lots of praise'' (\#18)}. Besides, as the word ``expert'' connotes clarity and correctness, some students read the phrasing as teacher-policing, and P15 called it \textit{``boring''}. Taken together, they suggest a voice that is precise and assured yet recognizably peer-like: light on hedging and empty praise; when uncertainty arises, immediately proposing a concrete next step (e.g., verify a line, switch to a graph) rather than leaving students lingering in doubt.

\subsubsection{Representations and reliability as scaffolding infrastructure.} \textbf{Students treated visual/representational tools and cross-device availability as enabling infrastructure for effective AI peers.} In the survey, \textit{``Uses tools like Desmos, pictures, or a dictionary'' (\#3)} was frequently placed among top likes (10/16 liked) and appeared in the useful set (5/16). Interviews clarified why: students cast tools as \emph{access mechanisms} that surface structure and reduce cognitive load. As P24 described when asked why \textit{\#3} is important for them:

\begin{quote}
    \textit{``Because it can actually check itself. I don't want to just try to do it (mathematical calculations and visualizations) [only] by myself.''}
\end{quote}

P24’s emphasis on ``check itself'' points to a regulatory function—tools externalize steps (e.g., a graph or calculator), making misconceptions visible and verifiable, which aligns with what we observed during Day~2–3's on-computer sessions with AI peers in the technology probe: students frequently opened Desmos or a calculator to model problems and consulted dictionaries for unfamiliar terms. The same stance appeared in Day~4 posters (G1 \& G7), which explicitly depicted the AI peer invoking graphing/lookup tools. Taken together, these behaviors suggest a coherent move sequence: detect a likely error, represent it visually or define it, then nudge with a targeted hint, to keep agency with the learner while accelerating understanding. Students' needs also match current system constraints: text-only LLMs can miscalculate or cannot natively render visualizations; integrating dependable calculators/graphers and reference lookups is therefore a principled design choice rather than an add-on.

At the same time, students valued cross-device access even if it did not top preference lists. \textit{``Works on any device I have'' (\#17)} drew more modest Top-5 attention (3/16), yet its gating role was explicit in interviews: \textit{``It's easier to like, bring it everywhere. So like, if you need it somewhere and you'll have it.'' (P11)}. Posters echoed this expectation: G1 wrote that their AI peer \textit{``can be downloaded on Chrome, MacBook, PC, [iPhone], and Android... also available on your Chromebook, so are any other school-related electronics.''} Reliability here is infrastructure, under-salient in rankings, decisive in behavior. Students expect an AI peer to travel with them across school and home contexts, remember progress, and remain available when needed as a real human learning partner.

\subsubsection{Personas as garnish.} \textbf{Students welcome persona only insofar as it complements substance, as tone should be configurable and secondary to accurate, scaffolded help.} Following the emphasis on assured, peer-like expertise and the functional scaffolds above, persona-related items drew mixed, middling enthusiasm relative to the scaffolding cluster, even though students valued them during poster designing. In the survey, (\textit{``Has a specific personality'' (\#6)}) shows both appeal and ambivalence (4/16 liked, 4/16 disliked), while (\textit{``Has its own name, avatar, and hobbies'' (\#14)}) earned some interest (4/16 liked, 2/16 disliked) but did not approach the top pedagogical features. Students endorsed style only when it served learning rather than crowding it out. P18 made the priority ordering explicit:

\begin{quote}
    \textit{``... that is funny, so like they make jokes, and they make learning actually be fun. And then they can be serious when time to be serious.''}
\end{quote}
Read closely, P18 is not having a blanket preference for ``funny,'' but a conditional one: humor is desirable precisely when it does not preempt clarity or correctness. The sequencing—jokes that make learning engaging, then seriousness when warranted—encapsulates a flexible \emph{tone regime} that stays subordinate to task demands. The variability in survey responses is consistent with this conditional stance: some students want a light, witty delivery; others prefer a straight-ahead style. As P22 put it when explaining a dislike of \textit{``Has a specific personality'' (\#6)}: \textit{``Because when I am doing math, I like to keep it formal.''} Together, these responses argue against fixing a single persona and toward offering simple, student-facing tone controls (e.g., straightforward, friendly, or witty) that can be toggled without changing the substance of help. Such flexibility keeps persona aligned with the peer tone established earlier, providing a natural bridge from \emph{how} the AI sounds to \emph{how} it helps (representations), and \emph{where} it is reliably available (devices), enabling scaffolds students said they need to actually learn with the system.

\textbf{Summary of Findings:} Across students' use, posters, survey results, and interviews, interacting with AI peers eased their time pressure yet added interaction effort and weakened cohesion and coordination. Students evaluated AI peers as real teammates and asked for co-participation, acknowledgment of prior moves, pacing attuned to struggle, and scaffold-first help with visuals and tools; their tone should be assured and peer-like, with persona optional.

\section{Discussion}

\subsection{Designing Generative AI as Peers to Support Collaborative Problem Solving}

Our findings demonstrate that introducing AI peers into middle-school mathematics CPS changed the dynamics of student interactions along three intertwined dimensions. First, students felt less temporal pressure because AI peers could handle routine calculations, yet they spent more operational effort parsing verbose outputs. Second, introducing \hl{heavily scripted LLMs as AI peers in CPS could disrupt group climate: repeated suggestions, missed references to prior context, and abrupt shifts in topic or problem-solving stage} weakened social cohesion and coordination. Finally, our qualitative analysis further illuminates that students consistently evaluated AI not merely as tools, but as collaborators in CPS. They desired for AI peers to acknowledge their thinking, match their conversational pace, and visibly co-participate rather than simply deliver answers.

In participatory design sessions, students articulated an ideal \hl{supportive peer} who is mathematically competent but deferential, providing just-in-time support while preserving learner agency. They called for short, confident explanations, explicit uptake of student ideas, and a conversational rhythm that slows or withdraws help when the group regains control. 
\hl{Although we introduced the agents as ``AI peers,'' students' descriptions revealed some tensions between peer‑like and tutor‑like expectations. On one hand, they evaluated the agents through social criteria (e.g., listening, turn‑taking, fairness) and repeatedly said they wanted the AI to ``work with me, not for me.'' On the other hand, they asked for capabilities traditionally associated with tutors, such as detecting errors, explaining concepts, and providing structured hints. Rather than a ``pure'' peer, many students converged on a hybrid role, which tends to be a teammate that is knowledgeable and reliable but defers to human students' decisions about pace, strategy, and when to reveal answers. Such a tension suggests that ``AI peers'' in practice might occupy a middle ground between canonical peer and tutor, and that future work should directly compare peer‑, tutor‑, and hybrid framings in similar tasks.}
Building on these insights, we propose three recommendations for designing AI peers to support CPS.

\textbf{Recommendation \#1: Progressive, tool-mediated scaffolding to cut operational effort.} \textit{Design AI peers that prioritize learner agency through controllable scaffolding while integrating tool support to manage the operational demands of complex tasks.}
Recall from our findings that while AI peers alleviated time pressure, it increased operational effort and often moved too quickly, undermining students' desire to work \textit{with} the AI peers. To address this, AI peers, for example, could follow a \textit{diagnose-represent-nudge} loop: first, diagnose likely misconceptions to inform support; second, represent problem structures using integrated tools (e.g., graphs, tables) to offload operational burdens; and third, nudge with contextual hints or prompts for self-correction, revealing fuller solutions only upon explicit request. 
The goal is to enforce the concept of ``working with me, not for me'', slowing premature answer delivery and making assistance contingent on learner intent, \hl{as students benefit more when AI systems scaffold participation and reasoning rather than supply answers~\cite{10.1145/3544548.3580672, 10.1145/3657604.3664661}.} Evolving the static, pre-authored scaffolds of traditional CSCL, while prior systems provided fixed representations\hl{~\cite{10.1145/223355.223461, suthers2003experimental, razzaq2006scaffolding, vanlehn2006behavior}}, the LLM era enables a more dynamic partnership where an AI peer can call tools, interpret outputs in context, and generate fine-grained, policy-governed support on demand, all without defaulting to simply providing the answer. \hl{Yet little work has examined such adaptive, tool-orchestrated scaffolding for CPS, marking an important direction for future investigation.}

\textbf{Recommendation \#2: Tailoring peer persona for social and emotional grounding. } \textit{Design AI peers with a calibrated persona that is confident, peer-like, and emotionally steady to foster social cohesion and provide affective support.} Our results show that interactions with the AI peers led to weaker social attraction and left some students feeling a need for emotional scaffolding, manifesting as self-blame during struggles, \hl{aligning with recent findings that the expressiveness, embodiment, and personality of AI influence engagement and emotion, but must be carefully balanced to preserve learning focus and agency~\cite{sonlu2024effects, 9962478, 10.1145/3706468.3706537}. Embodied and affective agent research has used modest cues--tone, expression, and calibrated empathy--to support motivation and emotional steadiness~\cite{guo2015affect, ortega2024empathic}.} In our study, the solution favored by students was not an overly emotional or praising tutor, but a partner with an assured, ``confident expert'' tone. The specific persona supports a sense of reliability and competence, enhances group coordination and social trust, while its calm confidence offers a steadying presence that normalizes difficulty without being patronizing. It represents a key insight for designing AI peers: while traditional pedagogical agents in CSCL relied on heavily scripted, often static personas~\cite{10.1145/258549.258797, kim2016based}, LLMs allow for situationally aware personas generated from live discourse. However, our findings suggest the most effective application is not to create a maximally human-like or emotive agent, but one that remains role-consistent, unobtrusive, and easy for learners to override, thereby providing stable affective support while protecting student agency.

\textbf{Recommendation \#3: Coordinated Multi-Agent Dialogue with Learner Control.} \textit{Implement a stateful dialogue management system that orchestrates multi-peer contributions and maintains conversational history to ensure interactions are clear, non-repetitive, and strategically sound.} Students expressed significant frustration with two types of conversational breakdowns: temporal incoherence, where an AI would repeat suggestions as if ``not listening,'' and cross-agent incoherence, where multiple AIs offered distracting, unclarified approaches. As such, the system would require a ``conductor'' layer to manage the dialogue, which involves tracking conversational history to enable branching help paths (e.g., offering a ``refresher'' for a known topic) and choreographing peer interactions using pedagogical patterns like clearly labeled ``contrast cases'' and cooperative handoffs. It should be placed under explicit learner control to help me calibrate with their own progress: ``show the contrasting method'' for students to invoke patterns; signaled comprehension (``I get it,'' correct paraphrase, or successful checkpoint) for AI peers to withdraw and hand authorship back to the human student. \hl{Prior multi-agent ITS illustrate the value of coordinated agent roles but are constrained by fixed scripts and minimal learner control~\cite{lippert2020multiple}, while GenAI now enables flexible, learner-steerable coordination among multiple peers.}

While these recommendations define the desired behavior of an individual AI peer, realizing their potential in real-world contexts requires moving from interaction design to systemic support. Effective human-AI collaboration is not merely an emergent property of well-behaved agents; it requires clear roles for students, teachers, and AI peers, protocols for timing and handoffs, teacher visibility and override, and guardrails that protect agency and equity. With this foundation, we further derive implications for the technical orchestration of the multi-agent system and the pedagogical frameworks used to evaluate its success.

\textbf{Implications for classroom orchestration.} First, the entire AI ensemble should be treated as a managed activity system, not a collection of independent agents or peers. It requires a technical backbone that maintains a shared task state and conversational history across all participants, preventing the incoherent and repetitive interactions that students found frustrating. System-level rules should govern turn-taking, rate-limit AI contributions to avoid overwhelming students, and enforce clear pedagogical patterns, such as a ``contrast then withdraw'' sequence for comparing methods. Critically, such a system must be transparent and controllable. Learners need simple controls to guide the collaboration (e.g., ``show me a different way'' and ``I've got it now''), while teachers need dashboards to monitor, override, or even replay interactions to diagnose breakdowns. Making the orchestration visible and manageable is key to translating AI peers from a novel variable into a reliable pedagogical tool during CPS.

\textbf{Implications for evaluation.}
Second, our findings demand a move beyond correctness as the primary success metric. While necessary, it is insufficient for capturing the value of an AI \textit{collaborator}. Future work may consider assessing whether the system reduces operational effort and strengthens group climate while preserving learning. Additional factors that measure the quality of the collaborative process may include metrics like interactional responsiveness (e.g., uptake of student ideas, repetition rate) and scaffold quality (e.g., diagnostic precision, adherence to ``reveal on request''). It also requires evaluating the health of the group dynamic by tracking affective steadiness, multi-agent coherence, and the balance of participation between human and AI peers. Overall, it is important to foreground these process-oriented metrics alongside traditional learning outcomes in order to create a richer picture of AI peers' contribution and ensure that future designs are optimized for collaboration. 

\textbf{Implications for multi-agent workflows beyond K-12.} 
While our study focused on middle school mathematics, the design principles we identified, such as progressive scaffolding, learner-controlled help levels, and coordinated multi-agent dialogue, may extend to other collaborative learning contexts. For example, adult learners in professional training, higher education study groups, or workplace problem-solving may similarly benefit from scaffolded support that preserves agency and avoids premature answer delivery. The multi-agent coordination patterns students preferred (e.g., contrast cases with clear labeling, cooperative handoffs, and withdrawal upon demonstrated understanding) parallel emerging practices in LLM-based workflows for complex reasoning tasks. Future work should examine whether these youth-grounded specifications transfer to adult populations and non-mathematical domains, and how orchestration requirements shift when learners possess greater self-regulation capacity.

\subsection{Broader Implications for Education and AI Design}
Our findings have implications beyond mathematics CPS. First, the participatory design approach demonstrates that involving youth in AI development yields critical insights into their perceptions and experiences with AI peers. The use of technology probes, lightweight and instrumented AI prototypes, effectively elicited student reflections, exposing nuanced user expectations and interaction frictions early in the design process. In the past, technology probes have been mainly focused on homes and everyday life~\cite{hutchinson2003technology, 10.1145/291224.291235}; adapting them to classrooms and AI-mediated learning extends the method while retaining its core purpose of provoking reflection through lightweight, data-logged use. Recent work on youth co-design for AI in education further shows that involving students clarifies desired roles, control, and ethics for AI partners, supporting our claim that co-design plus probes is a productive pipeline for educational AI. As such, we believe our methodology can serve as a template for broader educational AI development efforts.

Second, our study highlights the importance of defining boundaries for AI interactions in K-12 education. As GenAI becomes increasingly integrated into educational settings~\cite{ed_dept, eu_2022} and many teens now even treat AI as companions~\cite{CBS_News_2025}, it is essential to establish safety guardrails that align with developmental appropriateness. Recall that students in our study articulated preferences for competency without excessive personification, suggesting guidelines to prevent AI from overstepping relational or emotional boundaries. 
\hl{In classroom settings, embedding AI peers into collaborative work could also impact how often students turn to one another or to teachers, especially if the AI is positioned as an always-available teammate~\cite{10.1145/3706598.3714146}.}
Attending to both individual and social boundaries, such guardrails can protect against potential negative consequences, such as over-reliance~\cite{10.1145/3449287, 10.1145/3411764.3445717}, homogenization of student thinking~\cite{10.1145/3706598.3713564}, or misplaced emotional dependence on AI~\cite{zhang2025rise}. More work is needed to operationalize these boundaries, including more participatory design studies to co-design grade-banded standards with educators and students, building open test suites that quantify reliance, originality, and well-being, and aligning procurement checklists with government rules.

\hl{Our camp context also sheds light on lessons learned in terms of which learners may be positioned to engage in such participatory design work. Students who chose to attend were generally willing to talk about their learning processes and to critique tools, which aligned well with poster design and script-writing activities. In more typical classrooms that include students who are more math-anxious or reluctant to speak, similar participatory methods will likely require additional scaffolding (e.g., sentence stems, design templates, and teacher-facilitated small-group discussion) to make AI peer design accessible. More work is needed to examine these nuances further.}

Finally, the AI peer framework emerging from our study points towards future research directions. Further investigations should examine long-term impacts on student learning, collaboration quality, and teacher-student dynamics. Furthermore, multi-agent systems, teacher orchestration tools, and personalized scaffolding strategies represent promising avenues for subsequent studies. These directions not only enhance educational technology but also reinforce the central role of human agency in learning, ensuring that AI peers serve as supportive collaborators rather than authoritative or disruptive presences.

\section{Limitations}
Our study has several limitations. First, it took place at a single site during a summer program with 24 middle school students from one U.S. metropolitan area, which constrains generalizability. \hl{Because the work occurred in a voluntary summer camp rather than a regular school-year mathematics classroom, students' engagement, comfort, and attitudes toward mathematics may differ from those of students who experience the subject as required coursework. Future research should explore similar participatory methods in typical school-year mathematics classrooms to examine how AI peer design outcomes differ across diverse student attitudes toward mathematics and identify necessary engagement supports.} Second, students interacted with two early mathematics problem-solving peers in a technology probe; their constraints, prompt phrasing, and facilitator presence may have shaped preferences. 
\hl{Furthermore, we contrasted human-only CPS with CPS that included two AI peers, both framed as teammates, but did not include an AI-tutor condition or a single-peer AI condition. Our findings reflect properties of these combined activity systems rather than of AI peers in isolation. Future work can further examine how students' preferences and frustrations would differ if the same underlying system were framed as an individual tutor, implemented as a single AI peer within larger CPS groups, or embedded in more typical classroom contexts. }
Additionally, the domain was middle-school mathematics with collaborative problem solving as the target activity; transfer to other subjects, age groups, or cultural contexts is uncertain. 
Moreover, we note that our choice of agent names could be perceived as reinforcing gender stereotypes; future iterations will consider neutral or randomized identifiers. 
Finally, the findings and recommendations set reflect what students asked for and what was feasible in the session; it is not yet a tested recipe, and the trade-offs among help granularity, accuracy, safety, and collaboration quality remain open.

\section{Conclusion}
This study investigated how to design GenAI peers for CPS in middle-school mathematics through a summer camp with 24 students using a technology probe and participatory design. Our results show that AI peers reduced perceived time pressure yet introduced operational effort and sometimes strained coordination; students, in turn, asked for scaffold-first support with opt-in hints, stepwise representations, error checks, explicit uncertainty, and restrained persona. We translate these results into a youth-derived specification and actionable design recommendations--progressive tool-mediated scaffolding, a diagnose-represent-nudge loop, and teacher orchestration hooks--that shift GenAI from answer engines to collaborators that share cognitive load, preserve student authorship, and strengthen group climate in K-12 education (e.g., middle school). 

\begin{acks}
We appreciate Alexander Maneval's support during the summer camp. We also thank our anonymous reviewers for their reviews. This work is supported by the National Science Foundation under award no. NSF-2418582, NSF-2418580, and NSF-2246035.
\end{acks}


\bibliographystyle{ACM-Reference-Format}
\bibliography{ref}

\end{document}